\documentclass[reprint,aps,pra]{revtex4-1}

\usepackage{dcolumn}
\usepackage{amsmath}
\usepackage{amssymb}
\usepackage{amstext}
\usepackage{bm}
\usepackage{array}
\usepackage{multirow}
\usepackage{color}
\usepackage{mathtools}
\usepackage{float}
\usepackage{graphicx}

\begin{document}


\title{Revealing the modal content of obstructed beams}

\author{Jonathan Pinnell}
\email{Jonathan.Pinnell@students.wits.ac.za}
\affiliation{School of Physics, University of the Witwatersrand, Johannesburg 2000, South Africa}
\author{Valeria Rodr\'{i}guez-Fajardo}
\affiliation{School of Physics, University of the Witwatersrand, Johannesburg 2000, South Africa}
\author{Andrew Forbes}
\affiliation{School of Physics, University of the Witwatersrand, Johannesburg 2000, South Africa}
\author{Saoussene Chabou}
\affiliation{Applied Optics Laboratory, IOMP Institute, University of Setif, Setif, 19000, Algeria}
\author{Karima Mihoubi}
\affiliation{Applied Optics Laboratory, IOMP Institute, University of Setif, Setif, 19000, Algeria}
\author{Abdelhalim Bencheikh}
\altaffiliation[Also at ]{Electromechanical Department, University of BBA, BBA 34000, Algeria}
\affiliation{Applied Optics Laboratory, IOMP Institute, University of Setif, Setif, 19000, Algeria}

\date{\today}

\begin{abstract}
In this work, we propose a predictor/indicator of the self-healing ability of coherent structured light beams: the field's modal content. Specifically, the fidelity between the obstructed and unobstructed beams' modal spectrum serves as a useful measure of the degree to which the beam will likely self-heal after an arbitrary obstacle. Since any optical field can be decomposed in terms of any chosen orthonormal basis, this analysis is, therefore, less restrictive than other methods for determining self-healing ability. Furthermore, since modal content is propagation invariant, this allows beam self-reconstruction to be studied in this way at any convenient transverse plane. As a case study, we present convincing experimental evidence for the superiority of the self-healing properties of Laguerre-Gaussian over Bessel-Gaussian beams; analysis that is facilitated primarily by the proposed measure.
\end{abstract}

\maketitle

\section{Introduction}
Self-healing (or self-reconstruction) of structured light fields has received a steady amount of interest in the optical community in recent years, starting with the seminal work in Ref. \cite{Bouchal1998self} and progressing to the numerous applications that have been fostered in fields such as imaging \cite{Fahrbach2012}, microscopy \cite{fahrbach2013self}, micromanipulation \cite{Mcgloin2003A}, communications \cite{li2017adaptive} and quantum key distribution \cite{Nape2018}. Self-healing describes the process whereby an optical field is able to reconstruct some intrinsic property (such as amplitude, phase, angular momentum etc.) after an obstacle. Initial work on the subject of self-healing showed the ability of non-diffracting beams \cite{Durnin1987}, such as Bessel beams, to self-reconstruct their intensity at some distance behind an obstacle \cite{Bouchal1998self}, but was later found to extend to polarisation \cite{chavez2004reconAM,Milione2015a} and orbital angular momentum (OAM) \cite{bouchal2002resistance}, including the self-healing of classically entangled \cite{otte2018recovery} and quantum entangled states \cite{mclaren2014self}. It was also shown that optical vortices have the ability to self-reconstruct \cite{vasnetsov2000selfOV}, although not necessarily about the same orbital axis. 

Typically, a geometric (ray) argument is given as the root for the self-healing behaviour \cite{Litvin2009} but a wave-optics description also exists \cite{aiello2014wave}. Contrary to the initial belief, many different families of beams possess a self-healing property \cite{arrizon2015selfhealing}; not just the non-diffracting type. The self-healing of Laguerre-Gaussian beams \cite{hernandez2019structured} and Hermite-Gaussian beams \cite{olivas2015hermite} has been demonstrated. More generally, it turns out that any beam has the capacity to self heal \cite{aiello2017unraveling}. It is known that the self-healing ability of the obstructed beam strongly depends on the form, size and position of the obstacle in a complex way \cite{litvin2013angular}. Note that the reconstruction is only perfect in the limit of a vanishingly small obstacle, otherwise, the field reconstruction is only partial with different beams possessing different self-healing abilities for different degrees of freedom. It was remarked that the obstacle affects the field's spatial frequency content, from which it follows that filtering theory should be a useful tool for the analysis of self-healing ability \cite{camara2011broken}; although this was not elaborated on. 
Various measures of the degree of self-healing have been proposed, but as yet none have been universally agreed upon. One such example requires the computation of various inner products over differently sized (and dynamically changing) domains \cite{arrizon2018modeling}. Another requires the determination of the full complex amplitude of the field at various propagation distances \cite{aiello2017unraveling}. Although a convincing argument can be made for the appropriateness of these measures, it would nevertheless be desirable to have a robust measure of self-healing which can be determined in an easier manner.

In this regard, we investigate here the self-healing properties of scalar optical modes via their modal content. We begin by providing a rationale for why modal content could be a useful predictor/indicator of an obstructed field's propensity to self-heal. We then demonstrate this usefulness experimentally in the context of comparing the self-healing ability of Laguerre-Gaussian (LG) and Bessel-Gaussian (BG) beams under similar circumstances. As a consequence, we confirm and extend previous theoretical work which conjectures that the self-healing properties of LG beams are more favourable to that of BG beams \cite{Mendoza-Hernandez:15}. This has important consequences for a wide range of applications where self-healing BG beams are currently being utilised, where LG beams may offer better performance.
\begin{figure*}[t]
    \centering
    \includegraphics[width=\textwidth]{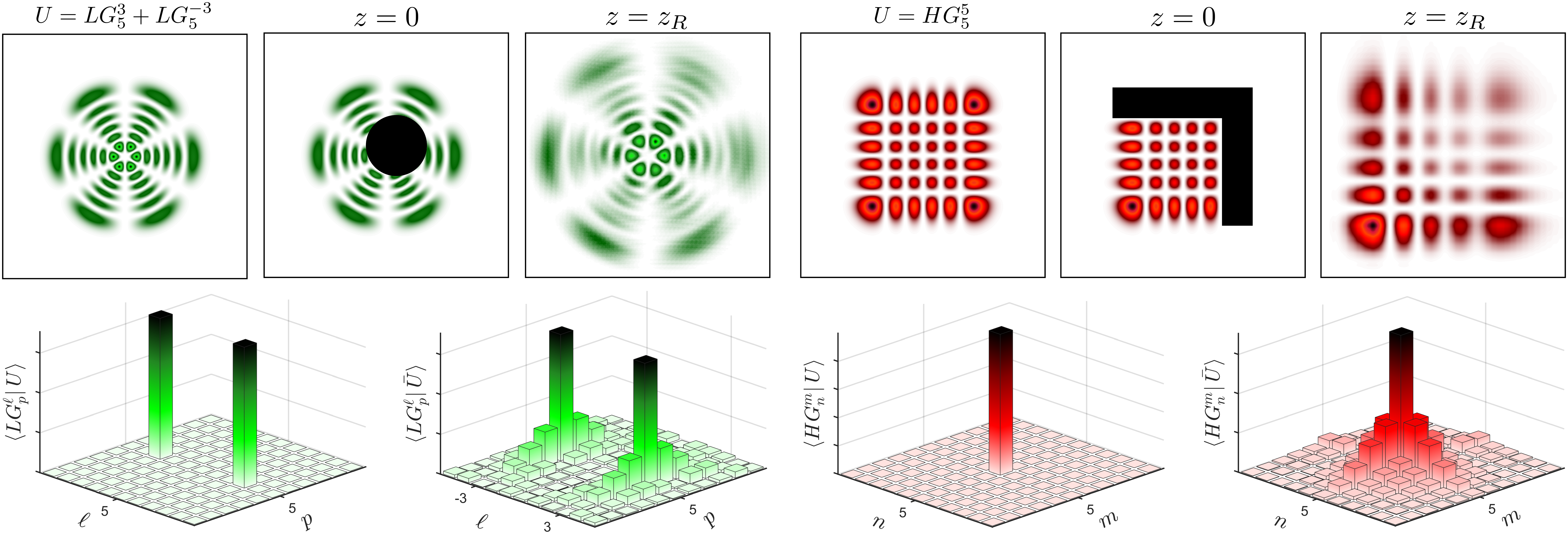}
    \caption{Simulation showing the modal content and propagated beam of an obstructed LG superposition and an obstructed HG mode, where the blacked-out regions correspond to the obstruction. For the LG (HG) beam, the field amplitude is maximally self-reconstructed by $94.8\%$ ($64.6\%$) at the plane $z = 0.6 z_R$ ($0.8 z_R$), with a mode spectrum fidelity of $F=0.81$ ($0.50$). For this particular example, there is a clear link between the mode spectrum fidelity and the degree of self healing.}
    \label{fig:Concept}
\end{figure*}

\section{Motivation}
Why would modal content be a useful indicator of self-healing ability? First, consider a scalar field $U(\mathbf{x}_\perp,z)$ propagating along the $z$ axis, where $\mathbf{x}_\perp$ are transverse spatial coordinates. Suppose that this field encounters an obstacle at $z = 0$ having transmission function $\mathcal{O}(\mathbf{x}_\perp)$. Now, any field can be expressed in terms of an arbitrarily selected orthonormal basis $\Phi_n(\mathbf{x}_\perp,z)$ where $n$ are the mode indices. Note that $n$ may represent multiple indices, for example the radial and azimuthal indices $(p,\ell)$, respectively, in the LG basis. The completeness of this basis ensures the validity of the following expansion,
\begin{equation} \label{eq:completeness}
    U(\mathbf{x}_\perp,z) = \sum_n c_n\, \Phi_n(\mathbf{x}_\perp,z) \,,
\end{equation}
where the set of numbers $c_n$ specifies the modal spectrum of the field in the chosen basis. Observe that the mode indices are constants and, crucially, are independent of the propagation coordinate $z$.

Now, there are numerous definitions of self-healing, but one such definition is that there exists some plane $z = z_{SH}$ (the so-called self-healing plane) where the obstructed field $\bar{U}(\cdot) = \mathcal{O}(\cdot) \, U(\cdot)$ has self-reconstructed and now closely resembles the original field,
\begin{equation}
    \bar{U}(\mathbf{x}_\perp,z_{SH}) \approx \alpha \, U(\mathbf{x}_\perp,z_{SH}) \,,
\end{equation}
where $\alpha$ is a constant that indicates how much of the beam power is reduced by the obstruction. Substitution of Eq.~\ref{eq:completeness} into the above and using the orthonormality of the basis functions yields
\begin{equation} \label{eq:SHMD}
    \bar{c}_n \approx \alpha \, c_n \,.
\end{equation}
This suggests that for self-healing to occur, the modal spectrum of the obstructed beam should resemble (to be quantified later) the modal spectrum of the unobstructed beam. Crucially, since the modal spectrum is invariant under propagation, this decomposition can be performed at any convenient plane after the obstruction. Further, since there exists an effective and simple optical procedure for finding the modal spectrum coefficients \cite{Flamm2012}, this allows the self-healing propensity of the field to be studied in a fully experimental manner, something which some previous analyses are incapable of doing. Moreover, this type of analysis is not restricted to a particular beam type since any laser field can be expanded into any desired modal basis.
\begin{figure*}[t]
    \centering
    \includegraphics[width=\textwidth]{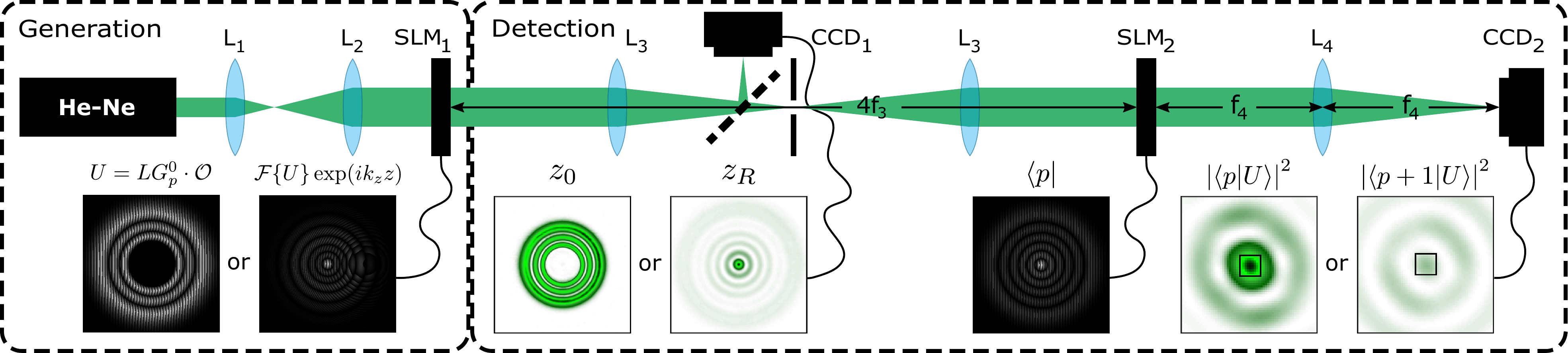}
    \caption{Experimental setup for switching between digital propagation and modal decomposition, enabled by flipping a pop-up mirror (dashed line). The images shown are for the case where a $LG_5^0$ mode is obstructed up to its third zero.}
    \label{fig:ExpSetup}
\end{figure*}

These ideas are highlighted qualitatively in a simulation shown in Fig.~\ref{fig:Concept} for two different structured light beams: a superposition of two LG modes and a single Hermite-Gaussian (HG) mode, each with different obstructions at different transverse positions at the waist plane (defined to be $z=0$). In each case, the obstructed fields were propagated by computing the Fresnel diffraction integrals over $z \in [0,2 z_R]$, where $z_R$ is the Rayleigh length, and the amplitude correlation (quantified in the next section) at each $z$ was computed. A modal decomposition in the respective basis was also performed before and after the obstruction. One can qualitatively observe that the LG superposition field experiences superior self-reconstruction which is correlated to the fidelity between the obstructed and unobstructed modal spectra.

Several intriguing aspects of self-healing are manifested here. We see, perhaps counter-intuitively, that the LG field has better self-reconstruction than the HG mode, even though the relative size of the obstruction is larger and more greatly reduces the initial beam power. This alludes to the fact that not only the size of the obstruction but its position is important. Since these fields are scaled-propagation invariant, a spatial frequency argument may be employed to partially explain the results: the HG obstruction more greatly impacts the high spatial frequency content which is where the bulk of the field information is stored \cite{camara2011broken}.

\section{Self-healing measures}
In what follows, we will study the self-healing of the field amplitude. We propose to compare the obstructed and unobstructed field amplitudes (obtained via a CCD camera) over a range of propagation distances $z$ according to a slightly altered and strengthened version of a measure used to determine the quality index of an original and test image \cite{wang2002universalImage}. One can view this measure as a kind of correlation measure between two matrices $x$ and $y$ having $N$ elements, which we define as,
\begin{align} \label{eq:Corr}
    C(x,y) = \frac{1}{2} &\left(1 - \frac{\sum_i (x_i-y_i)^2 }{N-1} \right) \left(1+ \frac{\sigma_{xy}}{\sigma_{x} \sigma_{y}} \right) \nonumber \\ \times &\left( \frac{2 \bar{x} \bar{y}}{(\bar{x})^2 + (\bar{y})^2} \right) \left( \frac{ 2 \sigma_{x} \sigma_{y}}{\sigma_{x}^2 + \sigma_{y}^2} \right) \,,
\end{align}
where,
\begin{align}
    \bar{a} &= \frac{1}{N} \sum_i a_i \,, \\
    \sigma_{a}^2 &= \frac{1}{N-1} \sum_i (a_i - \bar{a})^2 \,, \\
    \sigma_{ab} &= \frac{1}{N-1} \sum_i (a_i - \bar{a})(b_i - \bar{b}) \,.
\end{align}
The proposed correlation measure $C(x,y)\in[0,1]$ is a product of the quantities: normalised mean square error, correlation coefficient, luminance and contrast.
\begin{figure*}[t]
    \centering
    \includegraphics[width=\textwidth]{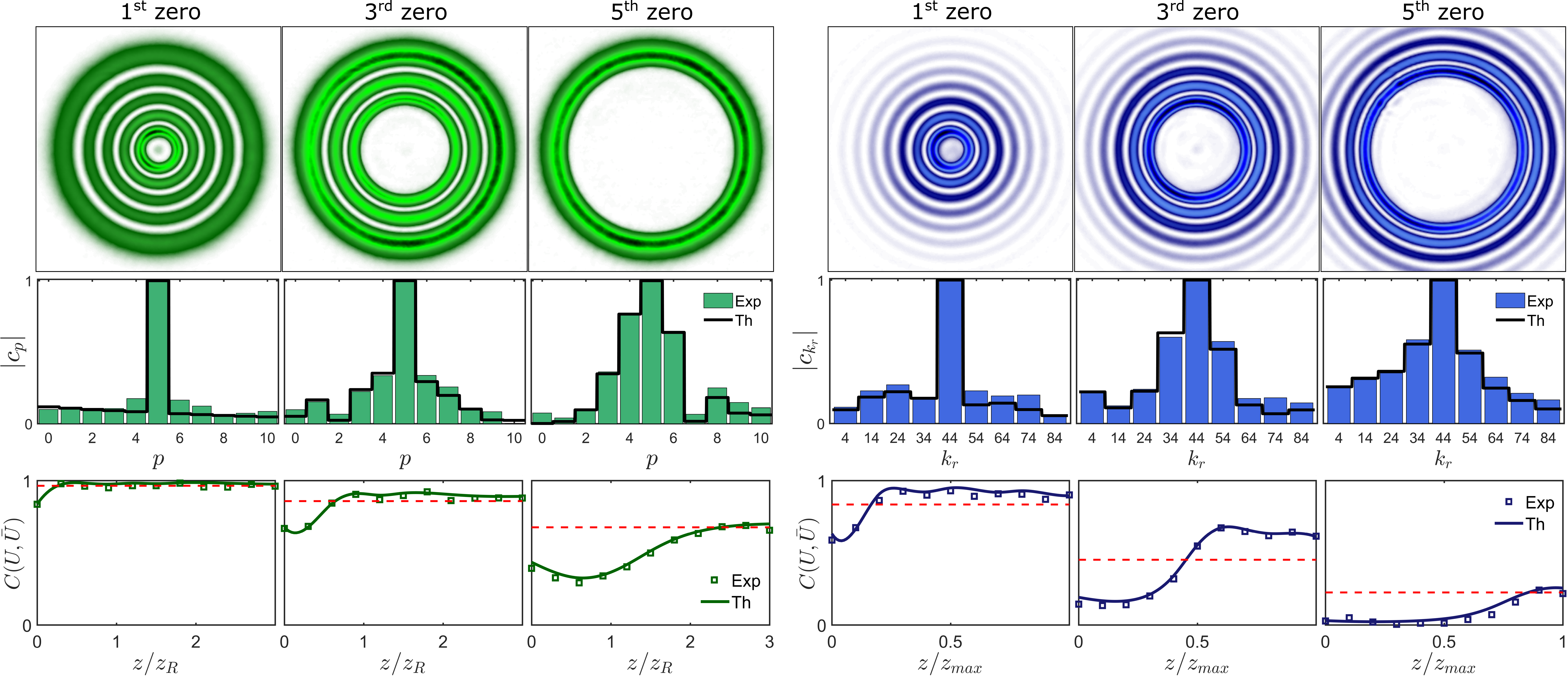}
    \caption{Experimental results of the LG versus BG showdown. Shown are the experimental beam images for different sized obstructions (first row), the corresponding rescaled radial mode spectra (second row) and the field amplitude correlation as a function of $z$ (last row). Dashed lines in the last row correspond to the modal spectrum fidelities with the unobstructed beam. Experiment (Exp) and theory (Th) are in excellent agreement.}
    \label{fig:ExpResults}
\end{figure*}

To measure the resemblance of the modal content, we compute the distance between the two modal spectra in terms of the fidelity,
\begin{equation}
    F(U,\bar{U}) = |\langle U|\bar{U} \rangle|^2 = \sum_n | c_n^* \, \bar{c}_n | \,.
\end{equation}
The above is an already widely utilised metric, such as in the determination of the fidelity of two pure quantum states \cite{Nielsen2002}.

\section{LG versus BG showdown}
We will study the modal content and self-healing of obstructed structured light fields in the context of comparing the self-healing ability of LG versus BG beams, as was done theoretically in \cite{Mendoza-Hernandez:15}. In doing so, we aim to experimentally confirm the theoretical prediction that the self-healing properties of LG beams are more favourable than that of the corresponding BG beams. Further, we show how the modal content can be effectively used to predict this behaviour.

In order to facilitate a fair comparison, it turns out that the BG and LG field profiles can be made similar if for any given $LG_p^\ell(\cdot)$ mode of Gaussian waist radius $w_0$, radial mode index $p$ and topological charge $\ell$, we choose for the BG beam,
\begin{align}
    k_r &= \frac{2 M}{w_0} \,, \label{eq:kr}\\
    w_{BG} &= M\, w_0 \,, \label{eq:wbg}
\end{align}
where $k_r,w_{BG}$ is the radial mode index and Gaussian waist radius of the BG mode and $M = \sqrt{2p + |\ell|+1}$ is the square root of the LG beam quality factor. This comparison follows from the asymptotic form of the associated Laguerre polynomials in the limit of a large radial index $p$ \cite{lebedev1972special}. In this regime, the Rayleigh range ($z_R$) of the LG beam and the maximum propagation distance ($z_{\text{max}}$) of the BG beam are identical, as can be seen by substituting the above into
\begin{align}
    z_R &= \frac{\pi w_0^2}{\lambda} \,, \\
    z_{\text{max}} &= \frac{2\pi w_{BG}}{\lambda k_r} \,.
\end{align}
The quantity $z_{\text{max}}$ is the distance over which the BG beam is quasi-nondiffracting; thereafter the beam energy expands to form a ring in the far-field. Similarly, LG beams are collimated within the depth of focus related to $z_R$, however, since they are scaled-propagation invariant they retain their amplitude profile in the far field.

\section{Experimental setup}
To test the ideas experimentally, we built the optical setup shown schematically in Fig.~\ref{fig:ExpSetup}. In actuality, this setup is a combination of two parts that each perform distinct tasks and can be alternated by flipping a pop-up mirror (denoted by the dashed line after lens $L_3$). When the pop-up mirror is flipped up, the setup can be used for digital propagation in accordance with \cite{schulze2012quality} which has the advantage of having no moving parts. When the pop-up mirror is down, the setup can be used for modal decomposition. In each case, the beam from a He-Ne laser is expanded and collimated to be approximately uniform over the first spatial light modulator (SLM).

In order to study the propagation dynamics of any particular optical field $U(\mathbf{x}_\perp,0)$, one can perform the following operations,
\begin{equation}
    U(\mathbf{x}_\perp,z) = \mathcal{F}^{-1} \{ \mathcal{F} \{ \, U(\mathbf{x}_\perp,0) \}\, \exp(i k_z z)\} \,,
\end{equation}
where $\mathcal{F}\{ \cdot \}$ denotes the Fourier transform and $k_z$ is the longitudinal component of the wave vector. This is equivalent to propagating the field via the angular spectrum approach. By encoding $\mathcal{F} \{ \, U(\mathbf{x}_\perp,0) \}\, \exp(i k_z z)$ on the first SLM and Fourier transforming this with a lens, the field at the focal plane is precisely the desired propagated field. By placing a CCD camera at this plane, the amplitude information of the propagated field can be acquired and the correlation between the obstructed and unobstructed amplitudes can be determined using Eq.~\ref{eq:Corr}. We then adjust $z$ digitally so as to enable the study of the field's propagation dynamics without any moving parts. Note that, in general, the optical field need not be known. The setup may be adjusted so that the SLM lies at the Fourier plane of another lens upon which the unknown field is impinged. In this case, only the propagation transfer function $\exp(i k_z z)$ is encoded. To save having to use another SLM, however, we choose to both generate the obstructed field and apply the propagation transfer function in a single step.

When the pop-up mirror is down, the modal decomposition part of the setup comprises the last three elements: SLM$_2$, lens $L_4$ and a CCD camera. This is known to be sufficient for optically determining the modal spectrum coefficients $c_n$ in Eq.~\ref{eq:completeness}. Specifically, the first SLM generates the obstructed beam $\bar{U}(\cdot) = \mathcal{O}(\cdot) \, U(\cdot)$ which is then imaged to the second SLM to be overlapped with the chosen set of basis functions $\Phi_n(\mathbf{x}_\perp,z)$. Since the mode spectrum is invariant with respect to $z$, for simplicity we chose to perform the optical inner product at the plane $z = 0$ (immediately after the obstacle). 

In the case of the LG basis, their inner product produces a Kronecker delta relation between the mode indices $(p,\ell)$ since these are discrete. For BG beams, however, the radial mode index $k_r$ is continuous and so one has to manually discretise this parameter to form an orthogonal set of modes. It was found that choosing the spacing as
\begin{equation}
    \Delta k_r > 1.5 \,\frac{2\pi w_{BG}}{\lambda f} \,,
\end{equation}
where $f$ is the focal length of the Fourier lens, ensures that the set of BG modes are orthogonal and can be used as a basis \cite{trichili2014detection}. For our optical system, a spacing of $\Delta k_r = 10\, \text{mm}^{-1}$ was selected.

A free parameter when choosing the basis functions in which to expand the field is the inherent scale parameter (such as $w_0$ in the case of Gaussian-type modes). It is important to choose an appropriate scale since selecting a sub-optimal scale will require extra radial modes in the expansion to reconstruct the field. In other words, a poor choice has the effect of increasing the spread of the radial mode spectrum. In this regard, a simple two-step approach has been offered to determine the beam size and propagation factor which is sufficient to extract the optimal scale of the modes in the modal decomposition (provided the family of modes which the beam belongs to is known) \cite{Schulze2012}. This enhancement improves the overall time efficiency of performing the modal decomposition and should be employed if the mode size is unknown.

In what follows, we consider circular obstacles centred on the optical axis and with a radius coinciding with the zeros of the field: the zeros of the associated Laguerre polynomials or the Bessel function for the LG and BG beams, respectively. Owing to the fact that the fields are comparable, due to Eqs.~\ref{eq:kr} and \ref{eq:wbg}, these zeros are approximately the same. One can view this obstacle function as effectively removing a certain number of inner rings from the beam. We also make the choice of using an LG mode with $p=5$ and $w_0 = 0.15\,\text{mm}$, which necessitates $k_r \approx 44\,\text{mm}^{-1}$ and $w_{BG}\approx 0.5\,\text{mm}$. For simplicity, we also restrict the experimental analysis to the case where $\ell = 0$, although we have found that the results extend analogously.

\begin{figure}[t!]
    \centering
    \includegraphics[width=\linewidth]{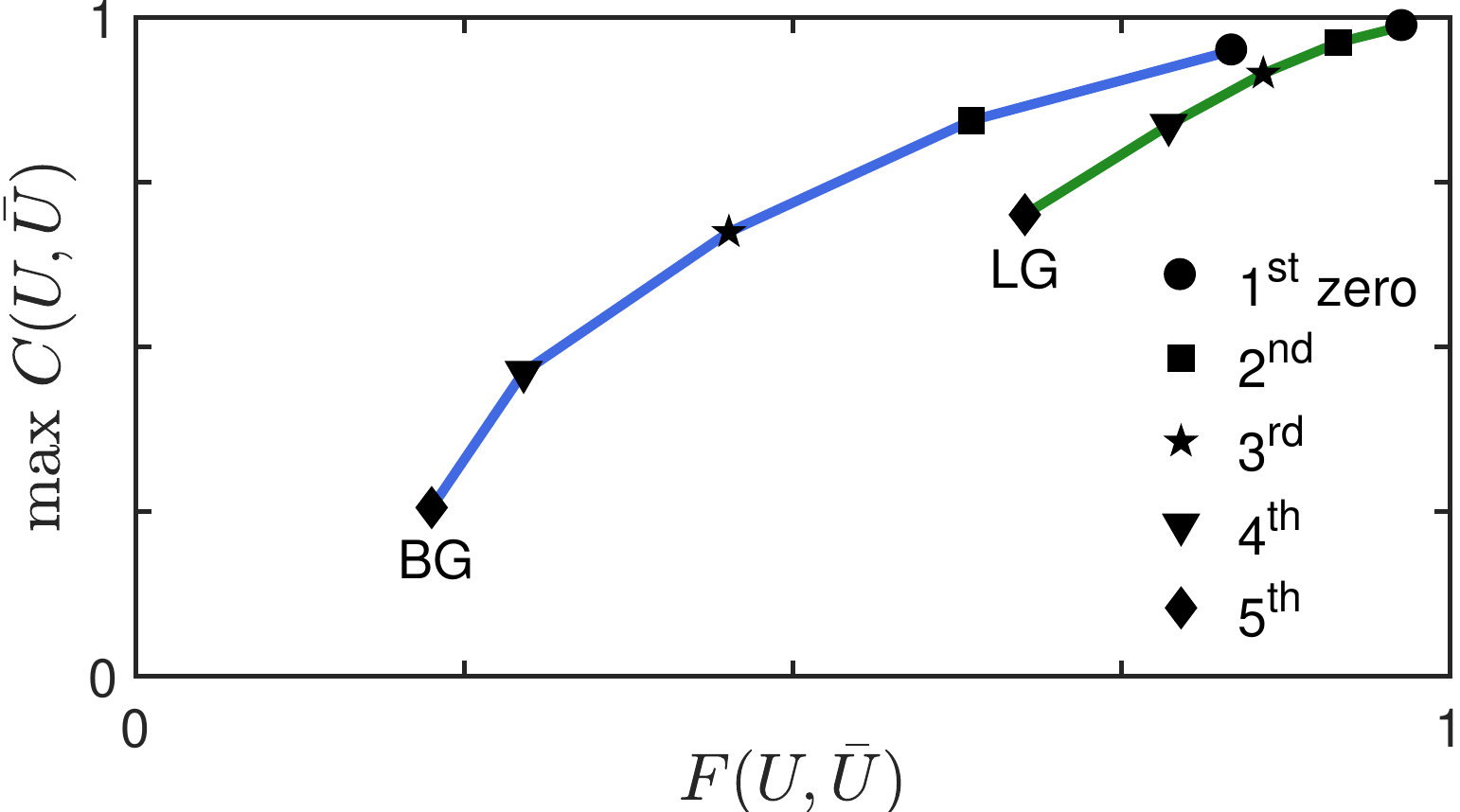}
    \caption{Maximum amplitude correlation versus modal spectrum fidelity for the LG versus BG showdown. Matching symbols correspond to the same obstruction.}
    \label{fig:CvsF}
\end{figure}
\section{Results}
A large subset of the experimental results are summarised in Fig.~\ref{fig:ExpResults} for three different obstruction sizes (the full results correspond to all five obstructions).  Self-healing is observed to some degree in all cases, as can be seen by the increase in the correlation coefficient after the obstruction at $z=0$. Most notably, the degree of self-healing, which we define as the maximum achieved correlation after the obstacle, is proportional to the fidelity of the obstructed and unobstructed mode spectra in every case. The mode fidelity in each case is denoted by the dashed horizontal lines in the last row of Fig.~\ref{fig:ExpResults}, all of which neatly follow along with the degree of self-healing.

We further note that the LG amplitude self-reconstruction persists after the self-healing plane (the plane where $C(U,\bar{U})$ is a maximum). This is in contrast to the BG amplitude whose correlation degrades to some degree after $z_{SH}$. Notice that the BG amplitude correlation is plotted over $z\in[0,z_{max}]$ whereas the LG amplitude correlation is plotted over three times this distance. This is facilitated by the fact that LG modes are scaled-propagation invariant, and so can be utilised over greater distances than the corresponding BG beams (which only propagate over a finite range). We also observe that for corresponding obstruction sizes, the modal content of LG beams is less disturbed and the maximum amplitude correlation is higher. 

A winner for the LG versus BG showdown is more clearly discerned from Fig.~\ref{fig:CvsF} where the maximum amplitude correlation is plotted against mode spectrum fidelity. Each of the five points correspond to each of the five similarly-sized obstructions (one for each zero of the field). We see that corresponding points in the LG line are higher and further to the right, indicating that for similar obstruction sizes the LG modes have superior self-healing and that the obstruction has a less damaging effect on the modal content. The proportionality between self-reconstruction and modal content is also clearly visible, further supporting the hypothesis that modal content can be a useful indicator of self-healing ability. Taken together, the facts presented above provide a very convincing argument for the superiority of the self-healing properties of LG beams over BG beams, with modal content being a useful indicator/predictor of this fact.

\section{Conclusion}
To summarise, we provided theoretical and experimental evidence indicating the utility of using a field's modal content to predict self-healing ability. As a case study, we applied these concepts in the context of a showdown between LG and BG beams, ultimately showing the superiority of the self-healing ability of LG beams. This has immediate consequences in applications where BG beams are currently being utilised where LG beams can offer better performance, such as the self-reconstruction of entanglement after an obstacle.



\bibliography{mypaperdatabase}

\end{document}